\RequirePackage{fix-cm}
\documentclass[twocolumn]{svjour3}          
\smartqed  
\usepackage{graphicx}
\usepackage{amsmath}
\usepackage{subfloat}
\usepackage{booktabs}
\usepackage[normalem]{ulem}
\usepackage[ruled,vlined]{algorithm2e}
\usepackage{epsfig}

\usepackage{placeins}
\usepackage{dblfloatfix}
\usepackage{afterpage}
\usepackage{subcaption}
\usepackage[colorinlistoftodos]{todonotes}
\usepackage{xcolor}
\begin{document}

\title{Modeling and Predicting Trust Dynamics in Human-Robot Teaming: A Bayesian Inference Approach 
}
\subtitle{ }


\author{Yaohui Guo        \and
        X. Jessie Yang 
}


\institute{Yaohui Guo \at
              Department of Industrial and Operations Engineering, University of Michigan, Ann Arbor. \\
              \email{yaohuig@umich.edu}           
           \and
           X. Jessie Yang (Correspondence author) \at
              Department of Industrial and Operations Engineering, University of Michigan, Ann Arbor. \\
            \email{xijyang@umich.edu}
}

\date{Received: date / Accepted: date}

\maketitle

\begin{abstract}
Trust in automation, or more recently trust in autonomy, has received extensive research attention in the past three decades. The majority of prior literature adopted a ``snapshot" view of trust and typically evaluated trust through questionnaires administered at the end of an experiment. This ``snapshot" view, however, does not acknowledge that trust is a dynamic variable that can strengthen or decay over time. To fill the research gap, the present study aims to model trust dynamics when a human interacts with a robotic agent over time. The underlying premise of the study is that by interacting with a robotic agent and observing its performance over time, a rational human agent will update his/her trust in the robotic agent accordingly. Based on this premise, we develop a personalized trust prediction model and learn its parameters using Bayesian inference. Our proposed model adheres to three properties of trust dynamics characterizing human agents' trust development process \textit{de facto} and thus guarantees high model explicability and generalizability. 
We tested the proposed method using an existing dataset involving 39 human participants interacting with four drones in a simulated surveillance mission. The proposed method obtained a Root Mean Square Error (RMSE) of 0.072, significantly outperforming existing prediction methods. Moreover, we identified three distinct types of trust dynamics, the Bayesian decision maker, the oscillator, and the disbeliever, respectively. 
This prediction model can be used for the design of individualized and adaptive technologies.


\vspace{10pt}
\end{abstract}

\section{Introduction} \label{Sec: Introduction}
The use of autonomous and robotic agents to assist humans is expanding rapidly. Robots have been developed for various application domains such as urban search and rescue (USAR) \cite{Murphy:2004ivba}, manufacturing \cite{Vaibhav2014}, and healthcare \cite{Rantanen2017}. For example, an in-home robot can be used to improve the coordination of patient communication with care providers and to assist the patient with medication management. 
In order for the human-robot team to interact effectively, the human should establish appropriate trust toward the robotic agents~\cite{Du2019,Hancock:2011eg,Lewis2018,DeVisser2020}.



Humans' trust in automation, or more recently trust in autonomy, has received extensive research attention in the past three decades. The diverse interest has generated multiple definitions of trust as a belief, attitude, and behavior \cite{Lewis2018}. In this paper, we use the definition by Lee and See \cite{See:2004vj}: Trust is the ``attitude that an agent will help achieve an individual's goals in situations characterized by uncertainty and vulnerability"  (see ~\cite{Lewis2018,Rossi2017,schaefer2014meta} for discussions on the definitions of trust and see \cite{Hoff:2015_HFJ,Merritt2013,Ullman2017} for examples using the Lee and See's definition). 

\begin{figure}[h!]
    \centering
    \includegraphics[width=0.8\linewidth]{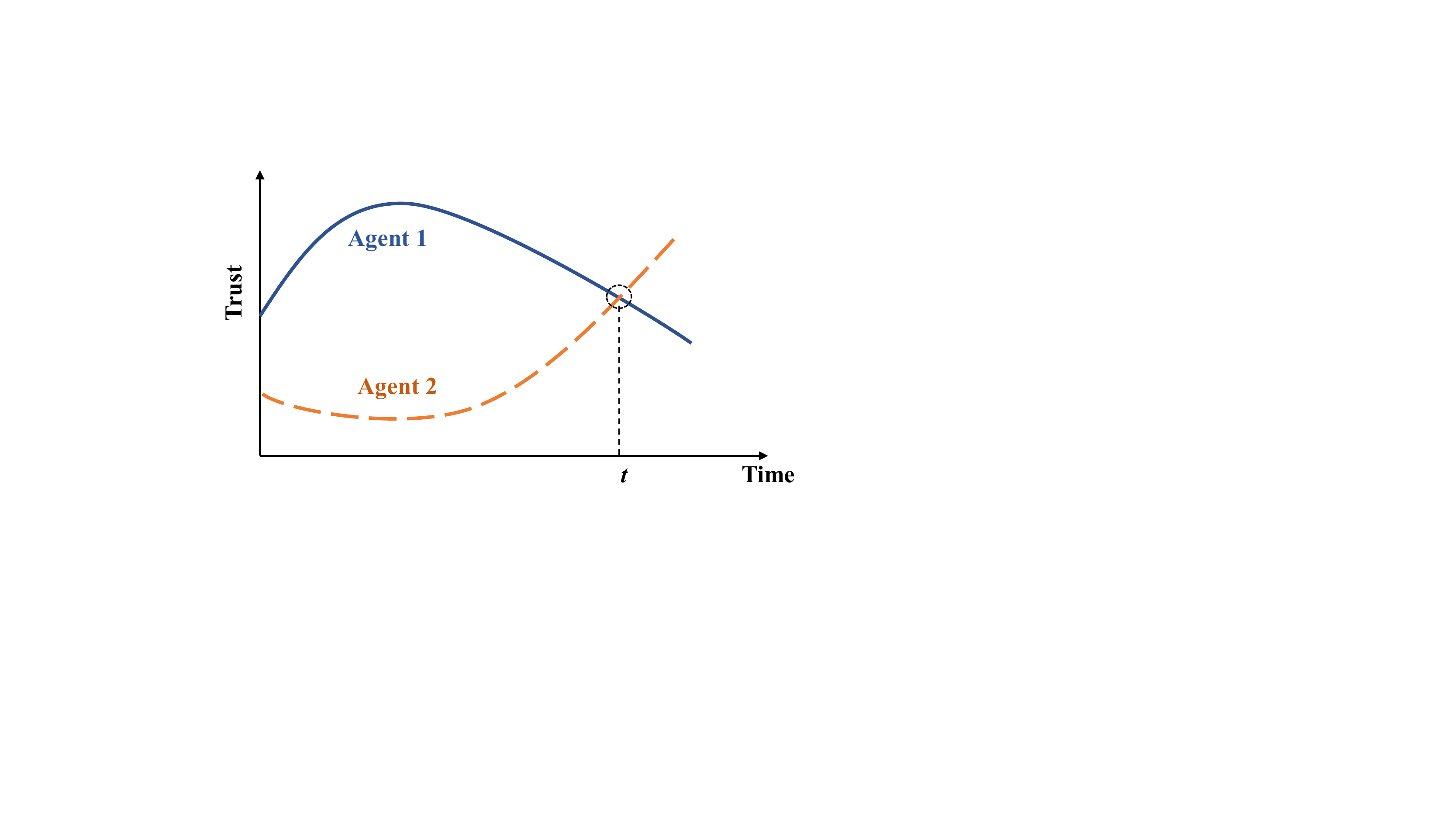}
     \caption{The static ``snapshot" view versus the dynamic view of trust. At time t, both agents have the same level of trust. However, their trust dynamics are different.}
     \label{fig:staticVSdynamic}
    \vspace{-10pt}
\end{figure}
Despite the research effort, existing research faces two major challenges. First, the majority of prior literature adopted a ``snapshot" view of trust and typically evaluated trust at one point, usually at the end of an experiment (Fig. \ref{fig:staticVSdynamic}). The static snapshot approach, however, does not fully acknowledge that trust is a dynamic variable that can strengthen and decline over time. With few exceptions (e.g. \cite{hu2016real,lee1992trust,Lu2020,Lu2019,manzey2012human,Xu:2015cc,Yang:2017:EEU:2909824.3020230,Yang:2016}), we have little understanding of a human agent's trust formation and evolution process after repeated interactions with a robotic agent \cite{DeVisser2020,Yang:2017:EEU:2909824.3020230}. Second, trust in automation is usually measured by questionnaires administered to the human agents. This approach introduces operational challenges, especially in high-workload and time-critical settings, because the human agent may not have the resource or time to report trust periodically. 

To address the two challenges, we develop a computational model that does not depend on repeatedly querying the human interacting with a robotic agent. Instead, this model infers a human's trust at any time by analyzing the robotic agent's performance history. 
We model a human agent's temporal trust using a Beta distribution and learn its parameters using Bayesian inference based on the history of the robotic agent's performance. This formulation adheres to three major properties of trust dynamics found in prior empirical studies: Trust at the present moment is significantly influenced by the trust at the previous moment~\cite{lee1992trust}; Negative experiences with autonomy usually have a greater influence on trust than positive experiences~\cite{manzey2012human,Yang:2016}; A human agent's trust will stabilize over repeated interactions with the same autonomous agent~\cite{Yang:2017:EEU:2909824.3020230}.
We test the proposed method using an existing dataset involving 39 human participants interacting with four drones in a simulated surveillance task. Results demonstrate that the proposed model significantly outperforms existing models \cite{lee1992trust,Xu:2015cc}. On top of its superior performance, the proposed model has another two significant advantages over existing trust inference models. As the proposed formulation adheres to human agents' trust formation and evolution process \textit{de facto}, it guarantees high model explicability and generalizability. Additionally, the proposed model is not based on the collection of human agents' physiological information, which could be difficult to collect. 


The remaining of the article is organized as follows. Section~\ref{sec:background} reviews the relevant literature on trust dynamics and prediction models. Section~\ref{sec:problem} formulates the trust prediction problem. Section \ref{sec:model} describes the proposed model and Section~\ref{sec:experiment} describes the dataset. Section \ref{sec:results} presents and discusses the prediction results of the proposed model. Section~\ref{sec:conclusion} concludes the study and suggests future research.


\section{Background} \label{sec:background}

As described in Section \ref{Sec: Introduction}, the majority of prior literature on trust in automation adopted a ``snapshot" view and typically evaluated trust at the end of an experiment. More than two dozen factors have been identified to influence one's ``snapshot" trust in automation. These factors can be broadly categorized into three groups: individual (i.e., the truster) factors, system (i.e., the trustee) factors and environmental factors.  Examples of individual factors are human's culture and age \cite{Rogers:2008,McBride:2011ix,Rau2009}. System factors include robot's reliability ~\cite{Salem2015,Wickens:2009False}, level of autonomy \cite{Rau2013}, adaptivity  \cite{Schneider2020} and transparency \cite{Wang:2016fl}, timing and magnitude of robotic errors~\cite{Desai:2013cmbaca,Rossi2017}, and robot’s physical presence \cite{Bainbridge2011}, vulnerability \cite{Martelaro2016}, and anthropomorphism \cite{Waytz2014}. Environmental factors include multi-tasking requirements~\cite{Zhang2017:HFES} and task emergency~\cite{Robinette:2016_HRI}. 

This ``snapshot" view, however, does not acknowledge that trust can strengthen or decay due to moment-to-moment interaction with autonomy. Only few studies emphasized the dynamic nature of trust and examined how trust changes as a human agent interacts with a robotic agent over time \cite{hu2016real,lee1992trust,Lu2020,Lu2019,manzey2012human,Xu:2015cc,Yang:2017:EEU:2909824.3020230,Yang:2016}. 

Manzey et al.  \cite{manzey2012human} noted two feedback loops in the human agent's trust adjustment process, namely a positive and a negative feedback loop. The positive loop is triggered by experiencing automation success, and the negative loop by experiencing automation failure. The negative feedback loop exerts a stronger influence on trust adjustment than the positive feedback loop \cite{lee1992trust,Yang:2016}. 
In addition, Lee and Moray \cite{lee1992trust} proposed an auto-regressive moving average vector (ARMAV) time series model of trust which calculated trust at the present moment $t$ as a function of trust at the previous moment $t-1$, task performance, and the occurrence of automation failures. Yang et al. \cite{Yang:2017:EEU:2909824.3020230} examined how trust in automation evolved as an \textit{average human agent} gained experience interacting with robotic agents. Results of their study showed that the average human agent's trust in automation stabilized over repeated interactions, and this process can be modeled using a first-order linear time-invariant dynamic system. The above-mentioned studies provide valuable insight into the trust dynamics of an average human agent. More recent studies used a data-driven approach to model trust dynamics. In this approach, trust is considered as information internal to the human that is not directly observable but can be inferred from other observable information \cite{Xu:2015cc}. For example, Hu et al.~\cite{hu2016real} proposed to predict trust as a dichotomy, i.e., trust/distrust, by analyzing the human agent's electroencephalography (EEG) and galvanic skin response (GSR) data. Similarly, Lu and Sarter \cite{Lu2019} proposed the use of eye-tracking metrics including fixation duration and scan path length to infer the human's real-time trust. Their follow-up study \cite{Lu2020} used three machine learning techniques, logistic regression, k-Nearest Neighbors (kNN), and random forest to classify the human's real-time trust level. Instead of using physiological signals, Xu and Dudek~\cite{Xu:2015cc} built an online probabilistic trust inference model based on the dynamic Bayesian network framework, treating the human agent's trust as a hidden variable which was estimated by analyzing the autonomy's performance and the human agent's behavior. In \cite{Xu:2015cc} the trust dynamics of each individual human agent was modeled. The above mentioned data-driven methods provided insights on how to predict a human's real-time trust by analyzing other observable information. However, they were subject to two limitations. First, some of them depend on using physiological sensors such as EEG and eye-tracking devices, which could be intrusive or be sensitive to noises \cite{hu2016real,Lu2019,Lu2020}. Second, as none of the existing models fully considered the empirical results showing how human agents adjust their trust \textit{de facto}, the resulting models could be limited in model explicability and generalizability.

\section{Problem Statement} \label{sec:problem}
In the present study, we aim to propose a personalized trust prediction model to predict each individual human agent's trust dynamics when s/he interacts with a robotic agent over time. In this section, we formulate the trust prediction problem mathematically. 


We consider a scenario where a robotic agent is going to work with a \textbf{\textit{new}} human agent on a series of tasks.
We denote the robot's performance on the $i^\text{th}$ task as $p_i\in \{0,1\}$, where $p_i=1$ indicates a success and $p_i=0$ indicates a failure. The reliability of the robotic agent, $r\in[0,1]$, is defined as the probability that the robot can succeed in the task. We assume that the robot has the same reliability while working with the new human agent.
At time $i$, after observing the robot's performance $p_i$, the new human agent will update his or her current trust $t_i \in [0,1]$ according to the robot's performance history $\{p_1,p_2,...,p_i\}$, where $t_i=1$ means the new human agent completely trusts the robotic agent and $t_i=0$ means s/he does not trust it at all. 

We assume that before the new human agent, the robotic agent has worked with $k$ other \textbf{\textit{old}} human agents,
and each of the old human agents finished $n$ tasks. Each old human agent reported his or her trust at the end of each task, so his or her trust history $T^j=\{t^j_1,...,t^j_n\}$ and the robot's performance history $P^j=\{p^j_1,...,p^j_n\}$ are fully available, $j=1,2,...,k$. 

\begin{figure}[h!]
    \centering
    \includegraphics[width=1\linewidth]{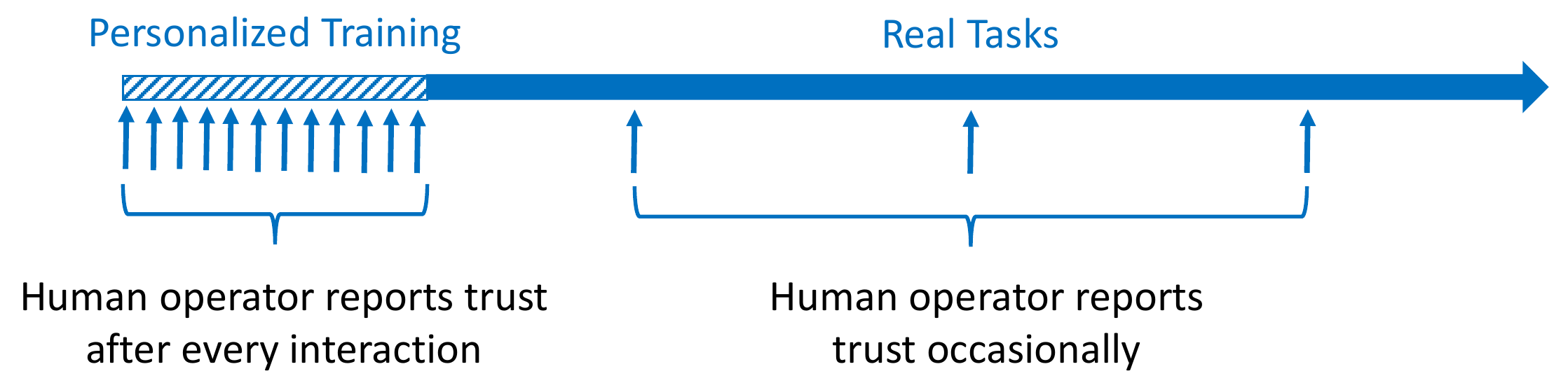}
     \caption{The new human agent receives a training session before performing the real tasks. During the training, the agent reports his or her trust after every interaction. When performing the real tasks, the agent reports his or her trust occasionally at their own discretion.}
     \label{fig:timeline}

\end{figure}

Before performing a real task, the new human agent receives a training session consisting of $l$ tasks (see Fig.  \ref{fig:timeline}). In the training session, the new human agent reports his or her trust after every task. After the training session, the new agent is to perform real tasks, during which s/he can choose whether to report his or her trust in the robotic agent at their own discretion.

The objective of the trust prediction problem is to predict the new human agent's trust $t_m$ after s/he finishes the $m^\text{th}$ task, based on the robot's performance history $P_m=\{p_i|i=1,2,3,...,m\}$, trust history during the training session $T_m^t=\{t_i|i=1,2,3,...,l\}$, occasionally reported trust $T_m^o=\{t_i|i\in O_m,O_m\subset \{l+1,l+2,...,m-1\}\}$, and the data $T^j$ and $P^j$ from the $k$ old agents, $j=1,2,...,k$. Here, $O_m$ is an indicator set: $O_{m}=O_{m-1}\cup \{m-1\}$ if the user choose to report his trust after the $m-1\text{th}$ task, otherwise $O_{m}=O_{m-1}$. We define trust history at time $m$ as $T_m=T_m^o\cup T_m^t$.

This formulation applies to any interaction scenarios wherein the human and the robotic agent are interacting with each other repeatedly and the human can observe the robotic agent's task performance over time. For example, a newly purchased in-home robot reminds an elderly adult of an upcoming monthly medical check-up. The elderly adult does not double-check the calendar and shows up at the doctor's office. Until then s/he finds out that the appointment has been re-scheduled by the doctor but the robot has not updated the calendar due to an error. Such a situation is considered a task failure by the robotic agent, and will most likely lead to a trust decrement. After the elderly adult interacts with the robot many times, s/he will probably have a more calibrated trust toward the robot and may not blindly follow the robot's monthly reminders anymore.

\section{Personalized Trust Prediction Model}\label{sec:model}
In this section, we summarize the major empirical findings on trust dynamics. After that, we introduce the proposed Beta distribution model and explain how it adheres to the empirical findings. Finally, we describe the Bayesian framework we use to infer the model's parameters.

\subsection{Major Empirical Findings on Trust Dynamics}
Based on the studies reviewed in Section \ref{sec:background}, a desired trust prediction model should adhere to three properties:

\begin{enumerate}
    \item Trust at the present moment $i$ is significantly influenced by trust at the previous moment $i-1$~\cite{lee1992trust}.
    \item \label{property:2} Negative experiences with autonomy usually have a greater influence on trust than positive experiences~\cite{manzey2012human,Yang:2016}.
    \item \label{property:3} A human agent's trust will stabilize over repeated interactions with the same autonomous agent~\cite{Yang:2017:EEU:2909824.3020230}.
\end{enumerate}

\subsection{Personalized Trust Prediction Model}
We use Beta distribution to model a human agent's temporal trust, for three reasons. First, Beta distribution, defined on the interval [0,1], is consistent with the bounded self-reported trust. Other distributions, e.g., Gaussian distribution,  could be unbounded. Second, Beta distribution fits the exploration-exploitation scheme and can be useful in a reinforcement learning scenario. Third, more importantly, the Beta distribution formulation adheres to the three properties in Section 4.1.

We use Bayesian inference to calculate the parameters defining the Bata distribution, because it provides better explainability compared to other machine learning methods, such as neural networks. Also, Bayesian inference provides a belief instead of point estimation of trust so it incorporates uncertainty. Moreover, Bayesian inference can leverage the population-wise prior for calculating model parameters for each individual human agent.



After the robotic agent completes the $i^\text{th}$ task, the human agent's temporal trust $t_{i}$ follows a Beta distribution:
\begin{equation}
t_{i} \sim Beta( \alpha _{i} ,\beta _{i})
\end{equation}
The predicted trust $\hat{t}_i$ is calculated by the mean of $t_i$
\begin{equation}
    \label{eq:predicted_trust}
   \hat{t}_i=E(t_i)=\frac{\alpha _{i}}{\alpha _{i} +\beta _{i}}
\end{equation}
$\alpha _{i}$ and $\beta _{i}$ are updated by
\begin{align}
\label{eq:trust_dynamics}
 \begin{array}{l}
\alpha _{i} =\begin{cases}
\alpha _{i-1} +w^{s} & ,\text{if} \ p_{i} =1\\
\alpha _{i-1} & ,\text{if} \ p_{i} =0
\end{cases}\\ \\
\beta _{i} =\begin{cases}
\beta _{i-1} +w^{f} & ,\text{if} \ p_{i} =0\\
\beta _{i-1} & ,\text{if} \ p_{i} =1
\end{cases}
\end{array}
\end{align}
where $p_i$ is the performance of the robot on the $i^\text{th}$ task. $\alpha_i$ and $\beta_i$ are the parameters of the Beta distribution and 
$w^{s}$ and $w^{f}$ are the gains due to the human agent's positive and negative experiences with the robotic agent. In other words, a success of the robot causes an increase in $\alpha_i$ by $w^s$ and a failure of the robot causes an increase in $\beta_i$ by $w^f$. The superscript $s$ stands for success and $f$ stands for failure. 

Next we explain how the model adheres to the three properties of trust dynamics. First, it is clear in Eq.~\eqref{eq:trust_dynamics} that the present trust is influenced by the previous trust, which satisfies the first property. Second, we calculate the difference between trust increment caused by the robot agent's success and trust decrement caused by the robot agent's failure at time $i$:
\begin{equation}
\begin{aligned}
\label{eq:diff}
&(\hat{t}_{i} |_{p_{i} =1} -\hat{t}_{i-1}) -(\hat{t}_{i-1} -\hat{t}_{i} |_{p_{i} =0})\\
=& \frac{1}{D}\left(\frac{w^{s} \beta _{i-1}}{D+w^{s}} -\frac{w^{f} \alpha _{i-1}}{D+w^{f}}\right)
\end{aligned}
\end{equation}
where $D=\alpha _{i-1} +\beta _{i-1}$.

\begin{figure}[h]
  \centering
  \includegraphics[width=0.6\linewidth]{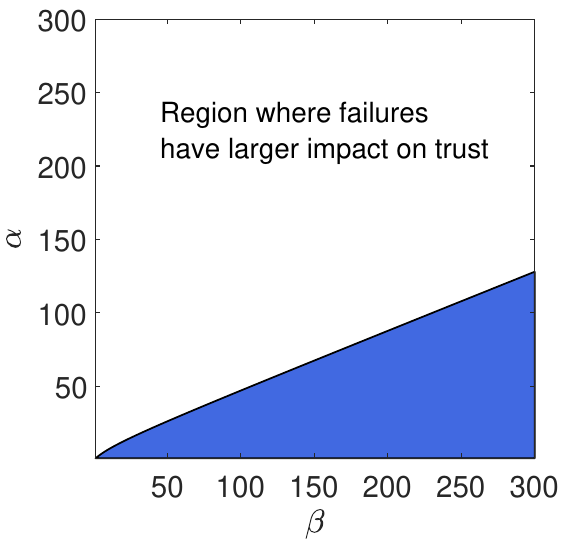}
  \caption{In the white region, the robot agent's failure would have a greater impact on trust than the robot agent's success. Here we set $w_f=50$ and $w_s=20$.}
  \label{fig:failureRegion}
\end{figure}

If $\alpha_{i-1}$ and $\beta _{i-1}$ are close, Eq.~\eqref{eq:diff} indicates that the robot agent's failure will lead to a greater trust change compared to the robot agent's success when $w^f>w^s$. More precisely, when 
$\frac{\alpha}{\beta} >\frac{w^{s} D+w^{s} w^{f}}{w^{f} D+w^{s} w^{f}}$, 
the robotic agent's failures will have a greater impact. An example is shown in Fig.~\ref{fig:failureRegion}. Within the white region the robot agent's failure would lead to a larger trust change. In Section 5 we show that $w^f>w^s$ is true for most human agents, such that the second property will be satisfied when the value of $w^s$ and $w^f$ are appropriately chosen. 

We assume the robot has a constant reliability $r$. After $n$ tasks, the robot accomplishes $n^s$ tasks and fails $n^f$ tasks. Then
\vspace{1mm}
\begin{equation}
    \label{eq:stabilization}
    t_i \sim Beta(\alpha_0+n^sw^s,\beta_0+n^fw^f)
\end{equation}
When $n\rightarrow \infty$, $t_n$ will be a point mass distribution centered at
\vspace{1mm}
\begin{equation}
    \frac{\alpha_0+n^sw^s}{\alpha_0+\beta_0+n^fw^f+n^sw^s}=\frac{rw^s}{rw^s+(1-r)w^f}
\end{equation}
which is a constant and it means trust stabilizes with repeated interactions. Therefore, the proposed model satisfies the three properties of trust dynamics.

To infer the model's parameters, after the $m^\text{th}$ trial, and given the robot's performance history $P_m=$

\noindent $\{p_1,p_2,...,p_m\}$, we determine trust $T_m=\{t_1,t_2,...,t_m\}$ by the parameter set
\vspace{1mm}
\begin{equation}
\displaystyle \theta =\left\{\alpha _{0} ,\beta _{0} ,w^{s} ,w^{f}\right\}
\end{equation}

Personalizing the trust model for the new human agent means finding the best $\theta$ for him or her. Here, we use the maximum a posteriori estimation (MAP) to estimate $\theta$, which is to maximize the posterior of $\theta$, given the robotic agent's performance $P_m$, trust history $T_m$ and robot reliability $r$. First, we have

\begin{align}
\begin{split}
& P( \theta \ |\ P_m,T_m,r)\\
 \propto & P( P_m,T_m,r\ |\ \theta ) \ P( \theta )\\
=&P( T_m\ |\ \theta ,P_m,r) \ P( P_m,r\ |\ \theta ) \ P( \theta )\\
  =&P( T_m\ |\ \theta ,P_m) \ P( P_m\ |\ r,\theta ) \ P( r\ |\ \theta ) \ P( \theta )\\
 =&P( T_m\ |\ \theta ,P_m) \ P( P_m\ |\ r) \ P( r) \ P( \theta )\\
  \propto &\prod _{t_{i} \in T_m} Beta( t_{i} ;\alpha _{i} ,\beta _{i}) \ \cdot P( \theta )
\end{split}
\end{align}


Then
\vspace{1mm}
\begin{align}
\begin{split}
    \theta = & \underset{\theta }{\text{argmax}} \ P( \theta \ |\ P_m,T_m,r)\\
=&\underset{\theta }{\text{argmax}} \prod _{t_{i} \in t} Beta( t_{i} ;\alpha _{i} ,\beta _{i}) \ \cdot P( \theta )\\
=&\underset{\theta }{\text{argmax}}\sum _{t_{i} \in T_m}\log( Beta( t_{i} ;\alpha _{i} ,\beta _{i})) \ +\log P( \theta )
\end{split}
\label{eq:theta_MAP}
\end{align}

The above equation shows that $\theta$ will be updated only when the human agent provides a new trust report. As $P(\theta)$ is unknown, the model needs to learn $P(\theta)$ first. This prior can be estimated by the empirical distribution of the parameters of the $k$ old human agents who have previously worked with the same robotic agent. The parameter $\theta_j$ of agent $j$ is estimated via the Maximum Likelihood Estimation (MLE):
\begin{equation}
\begin{aligned}
\theta_j  & =\underset{\theta }{\text{argmax}} \ P( T^j\ |\ \theta,P^j )\\
 & =\underset{\theta }{\text{argmax}} \ \prod\limits ^{n}_{i=1} Beta( t^j_{i} ;\alpha^j _{i} ,\beta^j _{i})
\end{aligned}
\label{eq:trainingMLE}
\end{equation}
where $\alpha^j_i$, $\beta^j_i$, $i=1,2,...$, are determined by Eq.~\eqref{eq:trust_dynamics}.

\section{Experiment and Dataset}\label{sec:experiment}
In this section, we describe the experiment and dataset used to test our proposed model. 

We use the dataset in Yang et al.~\cite{Yang:2017:EEU:2909824.3020230}. Participants in the study had an average age of 24.3 years (SD = 5.0 years) with normal or corrected-to-normal vision and without reported color vision deficiency. 

\begin{figure}[h!]
  \centering
  \includegraphics[width=1\linewidth]{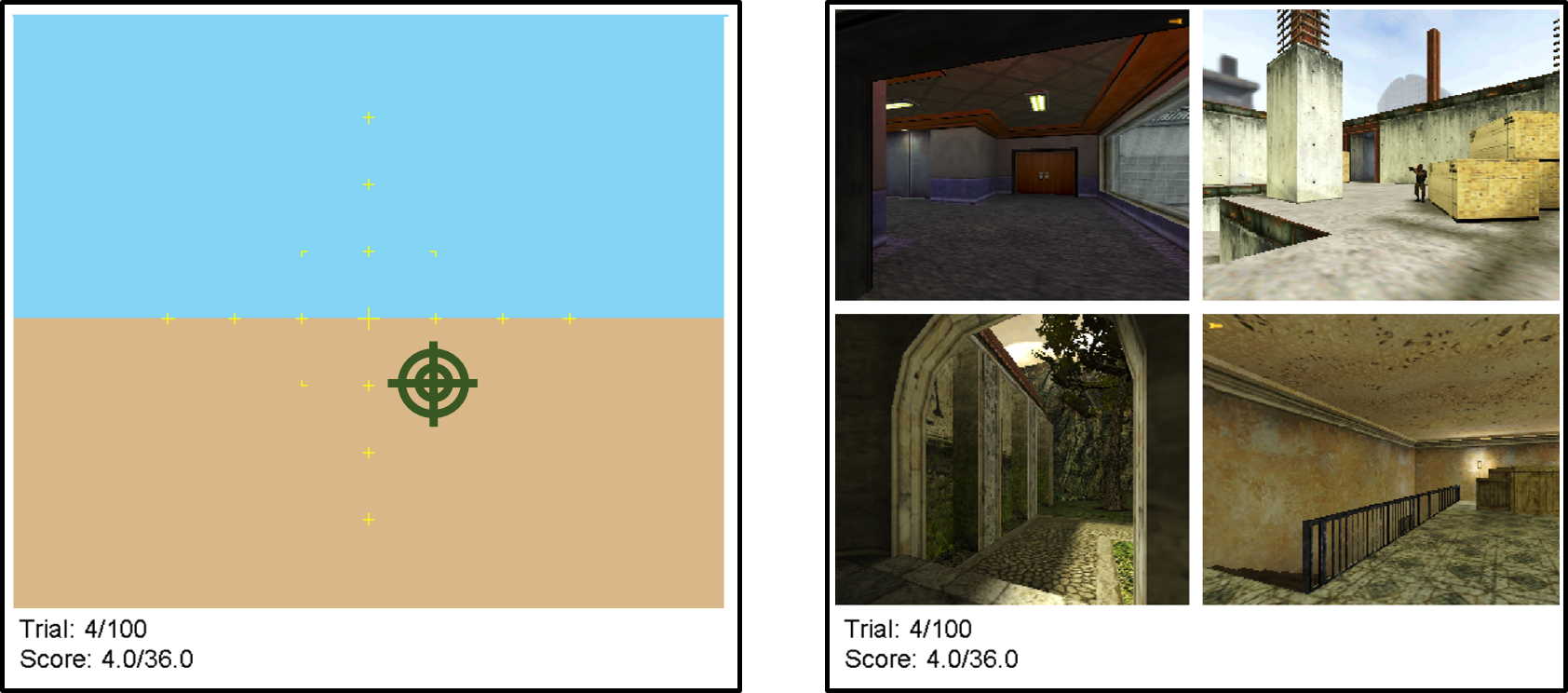}
  \caption{Dual-task environment in the simulation testbed. The two images show displays from the simulation testbed for the tracking (left) and detection (right) tasks respectively. Participants could access only one of the two displays at a time, and could switch between them.}
  \label{fig:dataEnv}
  \vspace{-2mm}
\end{figure}

All participants performed a simulated surveillance task with four drones. Each participant performed two tasks simultaneously (Fig.~\ref{fig:dataEnv}): controlling four drones using a joystick and detecting potential threats in the images captured by the drones. The participant was able to access only one task at any time and had to switch between the controlling and the detection tasks.

The drones were able to detect potential threats. They would report `danger' when a threat was detected. Due to environmental noises, the threat detection was imperfect. The system reliability of the drones was set as 70\%, 80\%, and 90\% according to the signal detection theory (SDT) \cite{Tanner1954,macmillan86creelman}. There were four states considering the drones' detection results and the true states of the world: hits, misses, false alarms, and correct rejections. As the drones cannot detect the threats perfectly, there is uncertainty involved in the task. For this particular experiment, a more contextualized definition of trust is a person's attitude that the drones will help him or her achieve his or her goal in the surveillance mission. 


The participants had two practice sessions to practice using a joystick. The two practice sessions consisted of a 30-trial block of the tracking task and an 8-trial block of both the tracking and the detection tasks. Hits, misses, false alarms, and correct rejections were illustrated during the 8-trial block. Then the participant completed the subsequent experimental block of 100 trials. The experiment lasted approximately 60 minutes with a 5-minute break at the halfway point. After each trial, participants reported their perceived reliability of the drones, trust in automation, and confidence. Each participant received compensation (a \$10 base) plus a bonus (up to \$5). The compensation scheme was determined from a pilot study, incentivizing participants to perform well. 

\section{Results and Discussion} \label{sec:results}

In the present study, we use data from the 39 participants who received binary detection alerts. We use the participants' self-reported trust and the drones' detection performance data according to the problem statement in Section~\ref{sec:problem}. To fully exploit the dataset, we use the leave-one-out method to evaluate the proposed model.
In each run, we select one participant as the \textbf{\textit{new}} human agent and consider the remaining 38 participants as the \textit{\textbf{old}} agents who previously worked with the drones. The trust history of the old agents and the robotic agent’s performance history are fully available for estimating $P(\theta)$; for the new human agent, we assume s/he performs $l$ trials during the personalized training session and thereafter when performing the real tasks s/he reports his or her trust every $q$ trials. In other words, after the new human agent's $m^{\text{th}}$ trial, where $m>l$, we predict his or her trust $t_m$ toward the robotic agent given his or her personalized training trust history $T_m^t=\{t_i|i=1,2,3,...,l\}$, the occasionally reported trust feedback $T_m^o=\{t_i|i=l+q,l+2q,l+3q,...,i<m\}$, as well as the data $T^j$ and $P^j$ from the old agents.



\subsection{Estimation of $P(\theta)$}

\begin{figure}[h]
  \centering
  \includegraphics[width=0.9 \columnwidth]{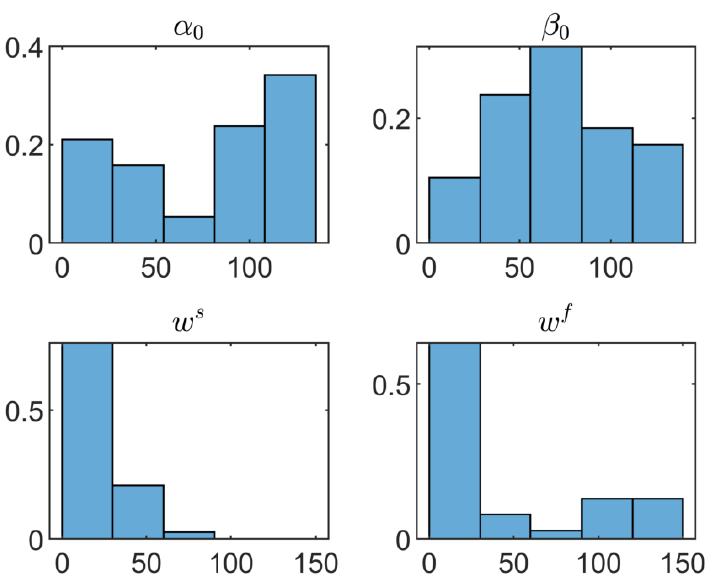}
  \captionof{figure}{Learned distribution of $w^s,w^f,\alpha_0,$ and $\beta_0$.}
  \label{fig:paraHist}
\end{figure}

We use Eq.~\ref{eq:trainingMLE} to estimate $P(\theta)$. 
Due to the small size of the dataset, we assume  $\alpha _{0} ,\beta _{0} ,w^{s} ,w^{f}$ are independent. We learn the prior distribution of the four parameters using MLE. Fig.~\ref{fig:paraHist} shows the empirical distributions of $\alpha_0,\beta_0,w_s,w_f$. Comparing the distributions of $\alpha_0$ and $\beta_0$ shows that $\alpha_0$ has a larger mean than $\beta_0$, which indicates that the participants in the experiment generally have a positive attitude toward the robotic agent. Comparing the distributions of $w^s$ and $w^f$ shows that in general $w^f>w^s$, which indicates most detection failures cause larger trust changes than detection successes.

\begin{figure*}[!t]
  \centering
  \includegraphics[width=\textwidth,height=0.45\textwidth]{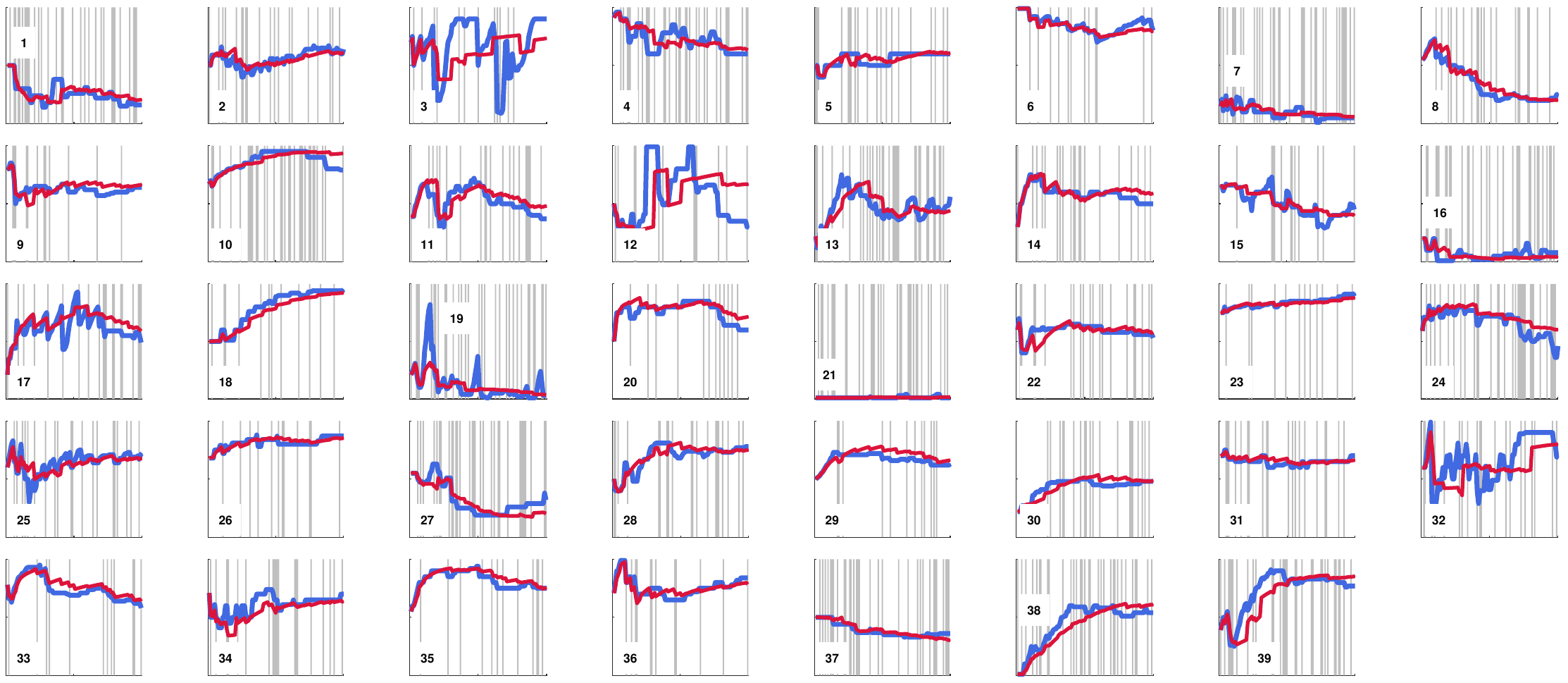}
  \caption{Trust prediction result for all participants under leave-one-out setting. X axis: trial number; Y axis: trust value. Blue curve: ground truth; red curve: predicted trust. The number in each plot is each participant's ID.}
  \label{fig:prediction_all}
\end{figure*}

\subsection{Prediction results and performance comparisons}
 Fig.~\ref{fig:prediction_all} shows the prediction results for all the 39 participants. The proposed model successfully captures the trust dynamics for many participants.
 
We compare the proposed model with two existing trust prediction models. We use root mean square error (RMSE) to evaluate the difference between the predicted value and the ground truth. The smaller the RMSE, the more accurate the prediction. 

The two models are the online probabilistic trust inference  (Optimo) model~\cite{Xu:2015cc} and the auto-regression moving average vector (ARMAV)~\cite{lee1992trust} model. We do not compare our model with \cite{hu2016real} or with \cite{Lu2020}, because our dataset lacks physiological data. Since the Optimo and the ARMAV models use different sets of variables, we modify them so all three models use the robot's performance history, but not other behavioral variables (e.g., human agent's intervention behaviors  \cite{Xu:2015cc}).

For each participant $h$, we calculate his or her RMSE using each prediction model $g$.

\begin{equation}
    \text{RMSE}_h^g =\sqrt{\frac{\sum ^{100}_{i=l+1}\left( {t_{i}} -\hat{t}_i^g\right)}{100-l}}
\end{equation}
where ${t_{i}}$ is the self-reported trust, $\hat{t}_{i}^g$ is the predicted trust calculated using method $g$ (i.e., our proposed model, ARMAV, and Optimo), 
and $l$ is the length of the personalized training session. 

The RMSE for each trust prediction model is calculated as the average of all the 39 participants:  $\text{RMSE}^g = \frac{1}{39}\sum_{h=1}^{39}{\text{RMSE}_h^g}$. Table \ref{tab:3methedComparison} details the mean and standard deviation of the RMSE values of the three models.

\begin{table}[h]
  \centering
  \caption{\label{tab:3methedComparison} mean and standard deviation (SD) of the RMSE values of the three models}
\begin{tabular}{lll}
\hline
                & \multicolumn{2}{c}{RMSE} \\ \cline{2-3} 
                & Mean        & SD         \\ \hline
Proposed method & 0.072       & 0.053      \\
ARMAV           & 0.101       & 0.052      \\
Optimo          & 0.139       & 0.080      \\ \hline
\end{tabular}

\end{table}

\begin{figure}
  \centering
  \includegraphics[width=0.8\columnwidth]{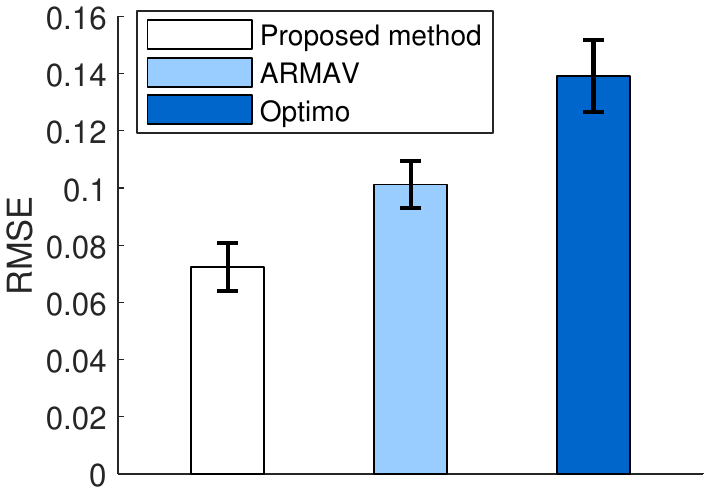}
  \captionof{figure}{Mean and standard error (SE) of RMSE for the three models. The error bar indicates the standard errors.}
  \label{fig:standardError}
\end{figure}

To compare the performance of the three trust prediction models, we conduct a repeated-measure Analysis of Variance (ANOVA), followed by pairwise comparisons with Bonferroni adjustments. The omnibus AN-OVA reveals a significant difference among the three models (F(2, 76)  = 21.64, $p <.001$). Pairwise comparisons reveals that our proposed model significantly outperforms ARMAV with a medium-large effect size ($t$(39) = 3.9, $p <.001$, Cohen's $d$ = 0.63), and Optimo with a large effect size ($t$(39) = 5.7, $p <.001$, Cohen's $d$ = 0.91). Fig.~\ref{fig:standardError} compares the three models.

The superior performance of our proposed model could have been due to two reasons: First, the proposed method captures the nonlinearity of trust dynamics, that trust stabilizes over repeated interaction with the same autonomous agent. In other words, the effect on trust due to a success or a  failure from the robotic agent changes as the interaction experience changes. While the first task failure from the robotic agent may cause trust to decline substantially, a robotic task failure after the human agent gains more experience may not. On the contrary, the ARMAV and Optimo models employ a linear rule for updating the predicted trust. It is clear in Fig.~\ref{fig:prediction_all} that most participants' trust varies at the start of the experiments and then stabilizes as more trials are completed. Second, although the three models define trust on a bounded interval [0,1], only our proposed method guarantees the predicted value to be bounded. The predicted trust value from ARMAV or Optimo needs to be truncated if it exceeds the defined boundary.

\subsection{Effects of trust report gap and training duration}
Since $w^s,w^f,\alpha_0,$ and $\beta_0$ are learned from the dataset, the only tunable parameters are the trust report gap $q$ and the number of personalized trials $l$. Thus it is necessary to understand the effects of the two parameters on the prediction results of our proposed method. 

\begin{figure}[!t]
  \centering
    \begin{subfigure}{0.7\columnwidth}
    \captionsetup{width=1.2\linewidth}
  \centering
        \includegraphics[width=1\columnwidth]{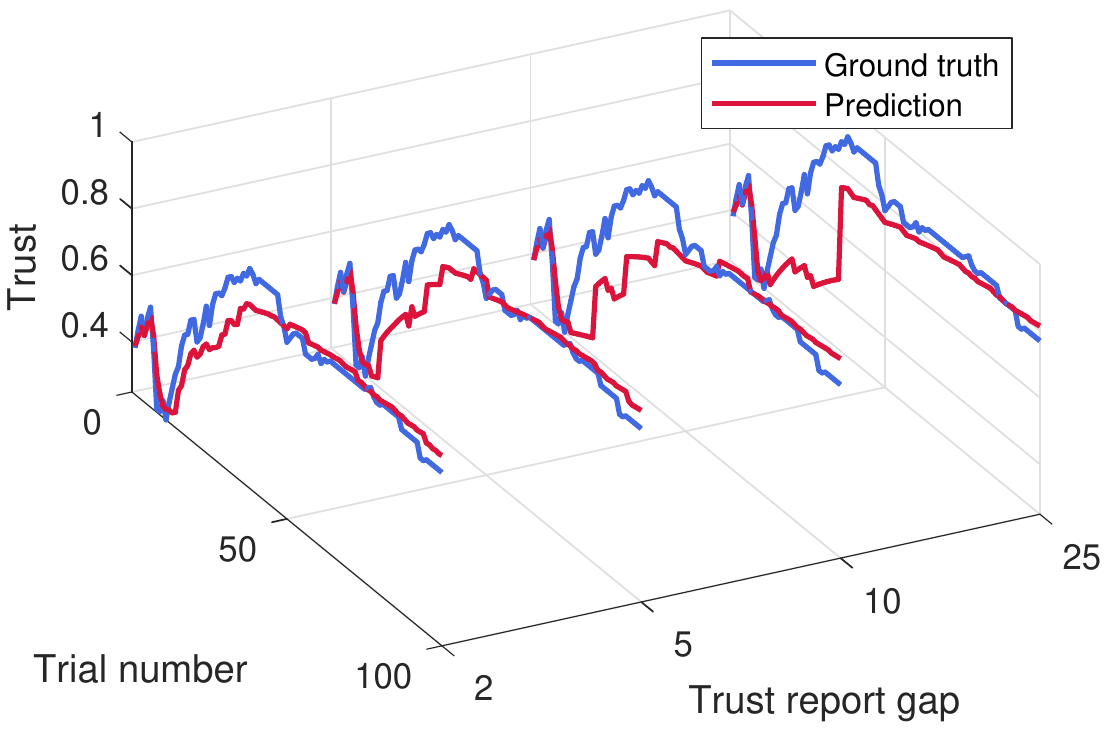}
        \caption{Prediction result with different trust report gaps. One participant's data is used in the figure.}
        \label{fig:different_gaps:dynamics}
    \end{subfigure}
    \begin{subfigure}{0.7\columnwidth}
    \captionsetup{width=1.2\linewidth}
  \centering
  \vspace{10pt}
        \includegraphics[width=1\columnwidth]{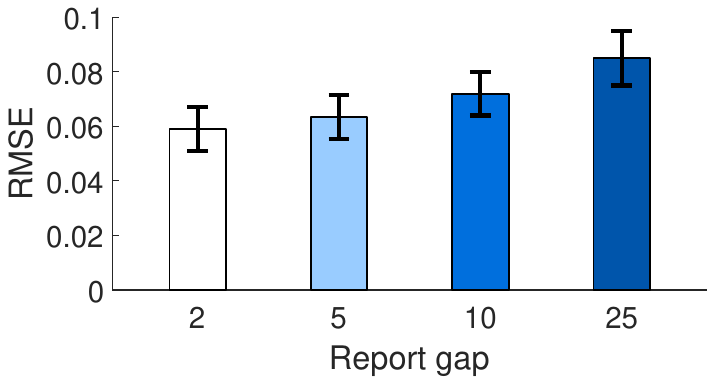}
        \caption{Mean and standard error (SE) of RMSE with different trust report gaps across all participants.}
        \label{fig:different_gaps:errorBar}
    \end{subfigure}  
    \caption{Prediction results with increasing trust report gaps $q$ from 2 to 25. Training duration $l$ is fixed at $10$.}
  \label{fig:different_gaps}
\end{figure}

To examine the effect of varying trust report gaps, we set the training duration $l=10$ and vary the trust report gap $q= 2,5,10$ and $25$. The average and SD of RMSE across the 39 participants are $0.059 \pm 0.050$, $0.064 \pm 0.052 $, $0.072 \pm 0.053$, and $0.085 \pm 0.062$ respectively. 
The effect of using different trust report gaps is illustrated further by using the data of one participant. Fig.~\ref{fig:different_gaps:dynamics} shows that as the trust report gap increases from 2 to 25, the deviance from the ground truth and the predicted values increases accordingly. Since the model parameters are updated when a new trust feedback is available, there are "jumps" on the prediction curve when the human agent chooses to report his or her trust after the training period. If the trust report gap is too wide, such as 25, the prediction accuracy is heavily harmed. This suggests that we need to carefully select the trust report gap such that the trust prediction accuracy can be maintained without disturbing the human agent extensively during real tasks.


\begin{figure}[!t]
  \centering
    \begin{subfigure}{0.7\columnwidth}
    \captionsetup{width=1.2\linewidth}
  \centering
        \includegraphics[width=1\linewidth]{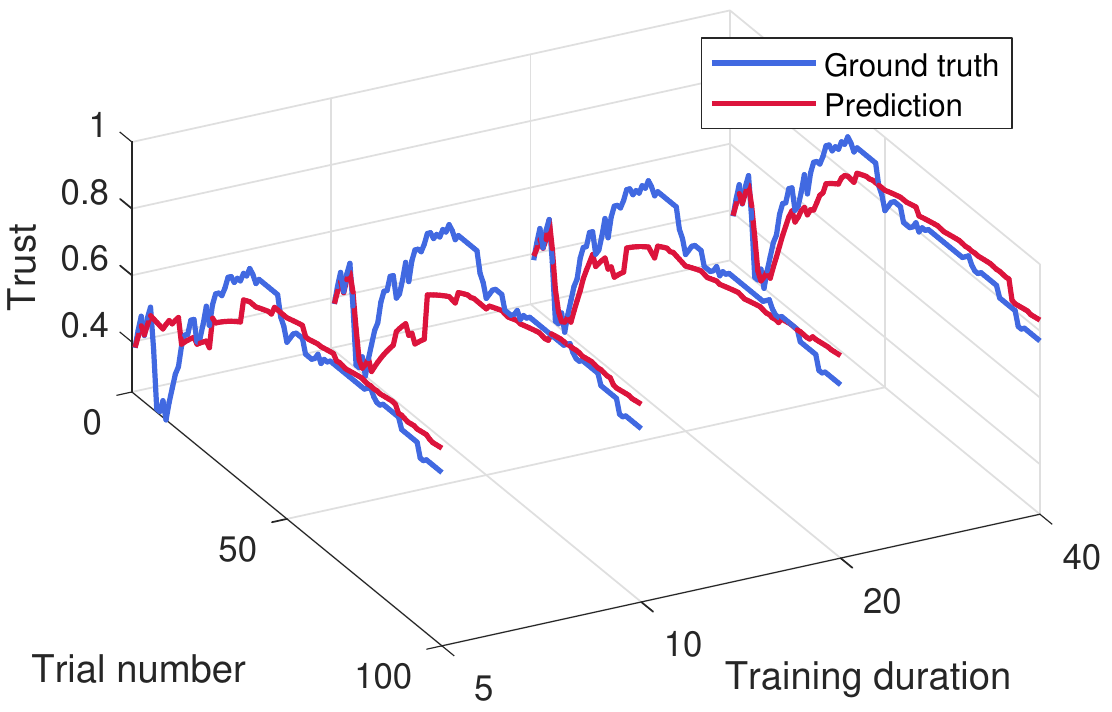}
        \caption{Prediction result with different training durations. One participant's data is used in the figure.}
        \label{fig:different_personalized_steps:dynamics}
    \end{subfigure}
    \begin{subfigure}{0.7\columnwidth}
    \captionsetup{width=1.2\linewidth}
  \centering
  \vspace{10pt}
        \includegraphics[width=1\linewidth]{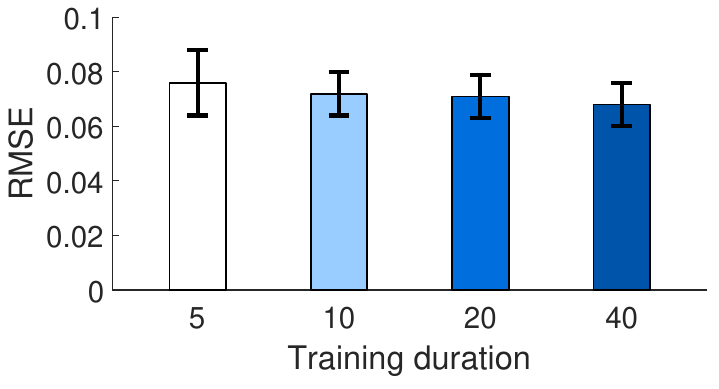}
        \caption{Mean and standard error (SE) of RMSE with different training durations across all participants.}
        \label{fig:different_personalized_steps:errorBar}
    \end{subfigure}  
    \caption{Prediction results with increasing training durations $l$ from 5 to 40. Trust report gap $q$ is fixed at 10.}
  \label{fig:different_personalized_steps}
\end{figure}





To examine the effect of using different training duration, we vary the 
training duration $l = 5, 10, 20, 40$ while fixing the trust report gap $q$ at 10. The average RMSE across the 39 participants $0.076 \pm 0.079$, $0.072 \pm 0.053$, $0.071 \pm 0.051$, and $0.068 \pm 0.052$ respectively (Fig. \ref{fig:different_personalized_steps}). 
This result suggests that the prediction error decreases with a longer personalized training session. 

\subsection{Three types of trust dynamics}

\begin{figure*}[!b]
  \centering
    \begin{subfigure}{0.3\textwidth}
  \centering
        \includegraphics[height=1.6in]{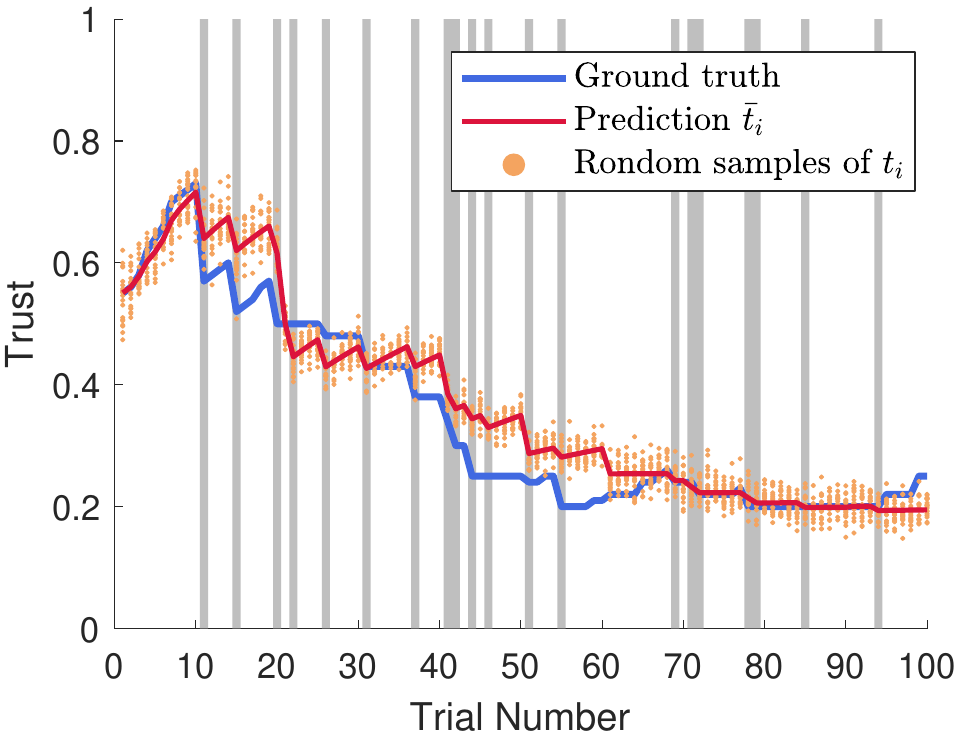}
        \caption{Bayesian decision maker}
        \label{fig:reasonable}
    \end{subfigure}
    \begin{subfigure}{0.3\textwidth}
  \centering
        \includegraphics[height=1.6in]{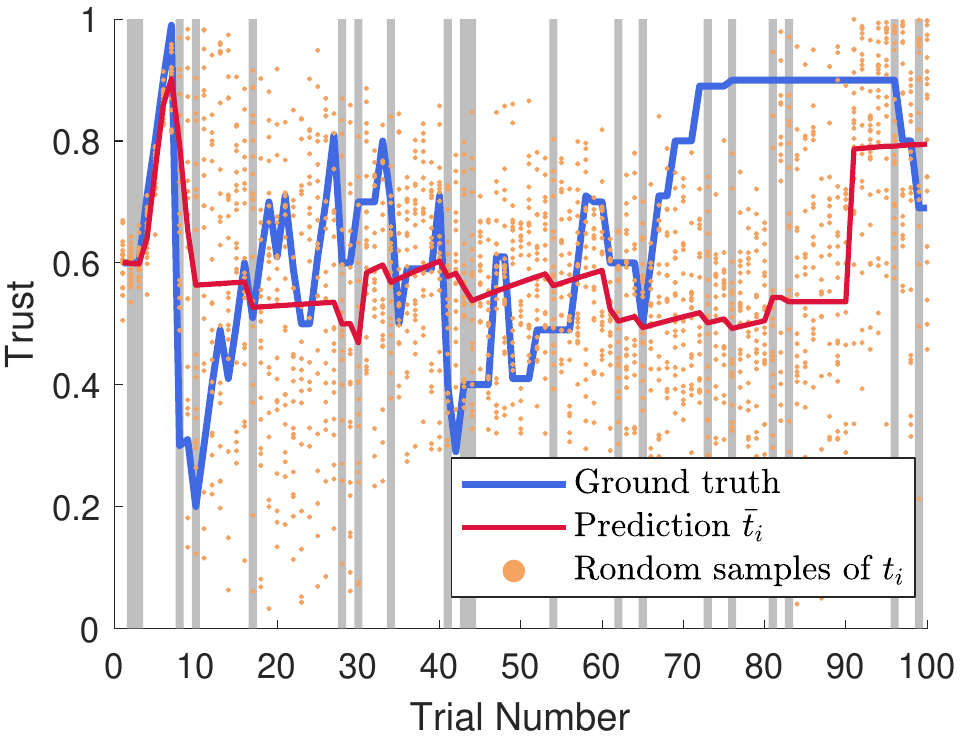}
        \caption{Oscillator}
        \label{fig:Oscillator}
    \end{subfigure}
    \begin{subfigure}{0.3\textwidth}
  \centering
        \includegraphics[height=1.6in]{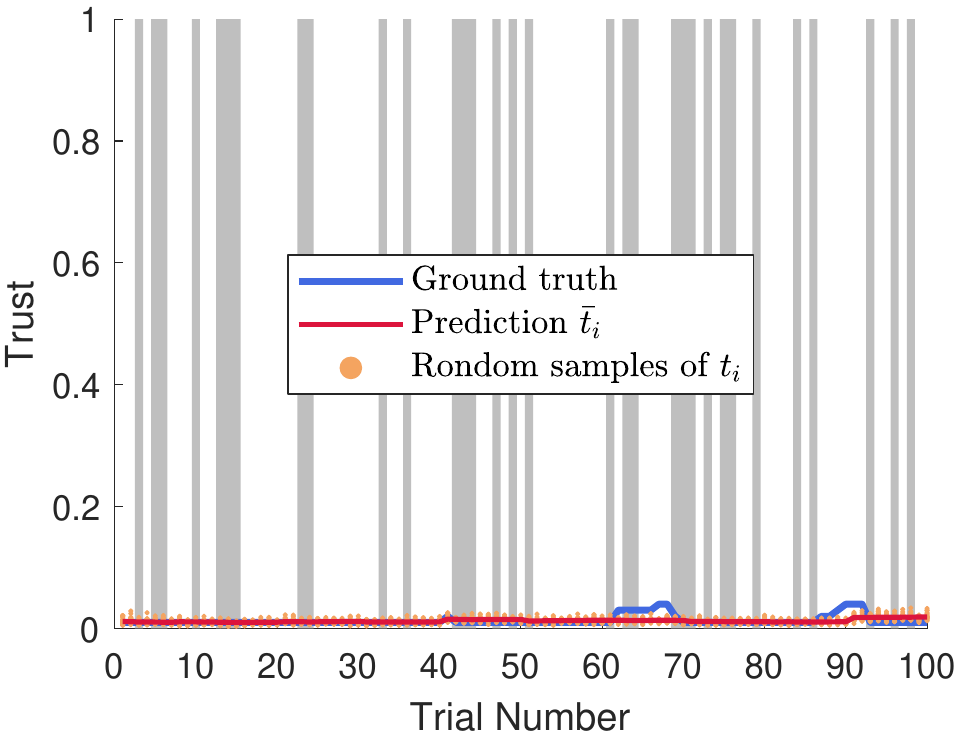}
        \caption{Disbeliever}
        \label{fig:untrustor}
    \end{subfigure}
    \caption{Three types of trust dynamics when human and autonomous agents interact over 100 interactions. $Trust$ at time $i$ is normalized ($trust_i \in [0,1]$).  Blue curve indicates the human agent's reported trust (ground truth). Red curve indicates the predicted trust.}
  \label{fig:catogories}
\end{figure*}

\begin{figure}[!t]
  \centering
    \begin{subfigure}{0.8\columnwidth}
    \captionsetup{width=1.2\linewidth}
  \centering
        \includegraphics[width=1\linewidth]{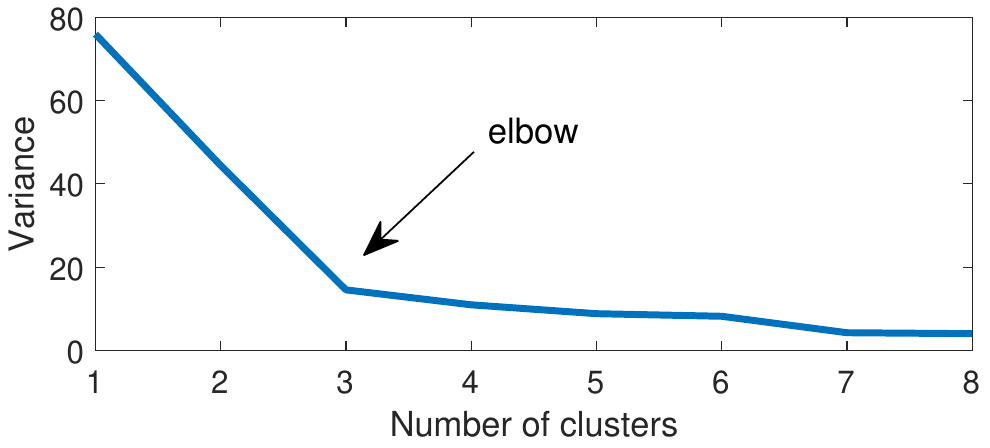}
        \caption{Based on the elbow, the best number of clusters is 3. The variance is the sum of point-to-centroid distances of all of the data points.}
        \label{fig:test2}
    \end{subfigure}
\\\vspace{2mm}
    \begin{subfigure}{0.8\columnwidth}
    \captionsetup{width=1.2\linewidth}
  \centering
  \vspace{10pt}
        \includegraphics[width=1\linewidth]{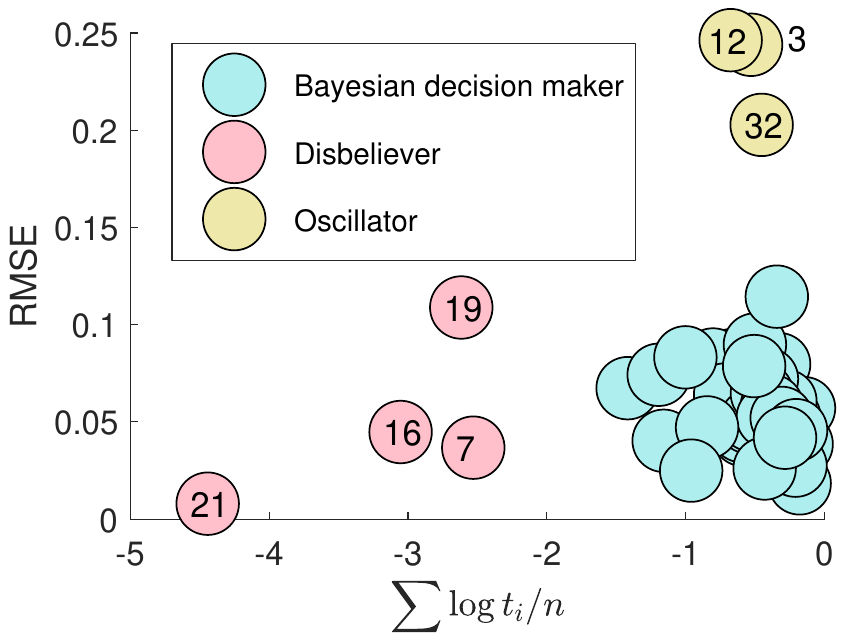}
        \caption{Clustering of participants. The two features are the RMSE and the average log trust. Examples of different clusters can be found in Fig.~\ref{fig:catogories}. Indices are participants' IDs shown in Fig.~\ref{fig:prediction_all}.}
        \label{fig:test1}
    \end{subfigure}  
    \hspace{0.2cm}
    \caption{Clustering of participants based on their trust dynamics.}
  \label{fig:clustering}
\end{figure}

Detailed investigation of Fig. \ref{fig:prediction_all} reveals the existence of different types of trust dynamics. To further investigate them, we perform k-means clustering \cite{Lloyd1982}. We find that while most participants' trust 
can be accurately predicted by the proposed method, some participants' self-reported trust significantly deviates from the predicted values. Moreover, four participants almost always reported very low trust in the experiment
(participants 7, 16, 19, 21 in Fig. \ref{fig:prediction_all}). Therefore, we select the RMSE and the average log trust as two features for the clustering analysis. RMSE measures how close the participant's trust dynamics follows the properties described in Section 4.1. Average log trust, defined by ${\sum_{i=1}^{n}\log t_i}/{n}$, can separate the participants who almost always report zero trust. We normalize the features across participants and determine the number of clusters by the elbow rule~\cite{thorndike1953belongs}, which is a commonly used heuristic to select the number of clusters. 
Fig.~\ref{fig:catogories} shows the three types trust dynamics and Fig. \ref{fig:clustering} shows the clustering analysis process. The first type is the Bayesian rational decision maker, shown in Fig.~\ref{fig:reasonable}. A Bayesian decision maker's trust dynamics follows the three properties that trust is dynamic, changes according to the robotic agent's performance, and stabilizes over repeated interactions. The second is the oscillator, shown in Fig.~\ref{fig:Oscillator}, whose temporal trust significantly fluctuates.
The third is the disbeliever, shown in Fig.~\ref{fig:untrustor}, whose trust in the robotic agent is constantly low.
The different types of trust dynamics may be related to each human agent's individual characteristics, such as their propensity to trust autonomy \cite{Merritt2013}.


\section{Conclusion}\label{sec:conclusion}


We proposed a personalized trust prediction model that adheres to three properties of trust dynamics characterizing human agents' trust development process \textit{de facto}. Trust was modeled by a Beta distribution with performance-induced parameters. The parametric model learned the prior of the parameters from a training dataset. When predicting the temporal trust of a new human agent, the model estimated the posterior of its parameters based on the interaction history between the human agent and the robotic
agent. The model was tested using an existing dataset and significantly outperformed existing models. On top of the superior performance, the proposed model has another two significant advantages over existing trust inference models. As the proposed formulation adheres to human agents' trust formation and evolution process \textit{de facto}, it guarantees high model explicability and generalizability. Additionally, the proposed model does not depend on the collection of human agents' physiological information, which could be intrusive and difficult to collect.

The proposed trust model complements the subjective measures of trust and can be applied to design adaptive robots. Accurately predicting trust in real time is the first step in designing robotic agents that can adapt to human agents' trust. For example, if a home companion robot detects an unexpected decline in trust by its human owner, the robot can adopt specific trust recovery strategies to regain the owner’s trust.


The results should be viewed in light of the following limitations: First, the proposed model assumes that the robotic agent's ability is constant across all the interactions. Second, it assumes the parameters are independent of each other. Third, the proposed model assumes that the robotic agent’s performance is dichotomous and immediately available after a task. Fourth, each participant in the experiment had 100 interaction episodes with the robotic agent in a relatively short period of time. To address the four limitations, further research is needed to test whether the proposed method would work for situations where a robotic agent learns and improves over time. The independence assumption can be removed once a larger dataset is available. Another promising future research direction is to examine how the proposed model should be modified for situations wherein the robotic agent's performance consists of multiple levels (e.g., extremely good, good, neutral, bad, extremely bad) or the agent's performance results are delayed. Further research is also needed to validate the proposed method with longer interaction episodes and to examine relationships between participants' individual characteristics and their trust dynamics.


\section{Acknowledgment}
This research was partially funded by ARL Cooperative Agreement Number W911NF-20-2-0087. The views and conclusions contained in this document are those of the authors and should not be interpreted as representing the official policies, either expressed or implied, of the Army Research Laboratory or the U.S. Government. The authors would also like to thank Drs. Chongjie Zhang and Cong Shi for useful discussions.

\section{Compliance with Ethical Standards}
Conflict of Interest: The authors declare that they have no conflict of interest.

\bibliographystyle{spmpsci}
\bibliography{Manuscript}

\end{document}